\documentclass[aps,prd,reprint,superscriptaddress,nofootinbib,floatfix,longbibliography]{revtex4-2}

\usepackage{amsmath,amssymb,bm,graphicx,hyperref}
\hypersetup{hidelinks}

\graphicspath{{figures/}}

\newcommand{\gagg}{g_{a\gamma\gamma}}
\newcommand{\sinc}{\operatorname{sinc}}
\newcommand{\dd}{\,\mathrm{d}}

\begin{document}

\title{Plasma-enabled magnet-free LSW search for eV-scale axions with laser-stimulated
regeneration}

\author{Xiangyan An}
\affiliation{State Key Laboratory of Dark Matter Physics, Tsung-Dao Lee Institute, Shanghai Jiao Tong University, Shanghai 201210, China}
\affiliation{State Key Laboratory of Dark Matter Physics, Key Laboratory for Laser Plasmas (MoE), School of Physics and Astronomy, Shanghai Jiao Tong University, Shanghai 200240, China}

\author{Min Chen}
\email{minchen@sjtu.edu.cn}
\affiliation{State Key Laboratory of Dark Matter Physics, Key Laboratory for Laser Plasmas (MoE), School of Physics and Astronomy, Shanghai Jiao Tong University, Shanghai 200240, China}

\author{Yipeng Wu}
\affiliation{State Key Laboratory of Dark Matter Physics, Tsung-Dao Lee Institute, Shanghai Jiao Tong University, Shanghai 201210, China}

\author{Boyuan Li}
\affiliation{State Key Laboratory of Dark Matter Physics, Key Laboratory for Laser Plasmas (MoE), School of Physics and Astronomy, Shanghai Jiao Tong University, Shanghai 200240, China}

\author{Wenchao Yan}
\affiliation{State Key Laboratory of Dark Matter Physics, Key Laboratory for Laser Plasmas (MoE), School of Physics and Astronomy, Shanghai Jiao Tong University, Shanghai 200240, China}

\author{Peng Yuan}
\affiliation{State Key Laboratory of Dark Matter Physics, Tsung-Dao Lee Institute, Shanghai Jiao Tong University, Shanghai 201210, China}

\author{Binbin Yan}
\affiliation{State Key Laboratory of Dark Matter Physics, Tsung-Dao Lee Institute, Shanghai Jiao Tong University, Shanghai 201210, China}

\author{Feng Liu}
\affiliation{State Key Laboratory of Dark Matter Physics, Key Laboratory for Laser Plasmas (MoE), School of Physics and Astronomy, Shanghai Jiao Tong University, Shanghai 200240, China}

\author{Zhengming Sheng}
\affiliation{State Key Laboratory of Dark Matter Physics, Tsung-Dao Lee Institute, Shanghai Jiao Tong University, Shanghai 201210, China}
\affiliation{State Key Laboratory of Dark Matter Physics, Key Laboratory for Laser Plasmas (MoE), School of Physics and Astronomy, Shanghai Jiao Tong University, Shanghai 200240, China}

\author{Jie Zhang}
\affiliation{State Key Laboratory of Dark Matter Physics, Tsung-Dao Lee Institute, Shanghai Jiao Tong University, Shanghai 201210, China}
\affiliation{State Key Laboratory of Dark Matter Physics, Key Laboratory for Laser Plasmas (MoE), School of Physics and Astronomy, Shanghai Jiao Tong University, Shanghai 200240, China}

\date{\today}

\begin{abstract}
In conventional light-shining-through-a-wall (LSW) experiments, strong magnetic fields mediate the conversion of photons into axions and their subsequent reconversion into photons. At eV-scale axion masses, however, photon--axion phase mismatch causes these conversions to lose coherence, substantially reducing the experimental sensitivity. We propose a magnet-free LSW scheme that addresses this limitation via plasma-based axion generation and laser-stimulated regeneration. At low laser intensity, a two-color dispersion-matched source generates axions in a narrow mass range set by the plasma dispersion. At relativistic laser intensity, a single-pulse plasma-broadened source generates axions over a broad mass range, as shown by particle-in-cell simulations. The Fourier spectrum of the source term $\mathbf{E}\cdot\mathbf{B}$ broadens around the laser branch, consistent with mixing between the laser and low-frequency plasma waves. For axion detection behind the wall, a counter-propagating laser stimulates axion decay into a detectable signal photon and a second photon emitted into the laser mode, with the laser frequency chosen to satisfy phase matching. For representative parameters, the requirement of at least one regenerated photon yields projected upper limits on the axion-photon coupling on the order $10^{-7}\,\mathrm{GeV}^{-1}$ over $m_a=0.1$--$1\,\mathrm{eV}$.
\end{abstract}

\maketitle

\section{Introduction}

Axions were introduced as a dynamical solution to the strong-$CP$ problem in quantum chromodynamics \cite{PecceiQuinn1977,PecceiQuinn1977b,Weinberg1978,Wilczek1978,Kim2010}. More generally, axionlike particles (ALPs) with a similar two-photon coupling arise in a broad class of extensions of the Standard Model \cite{Svrcek2006,Gelmini1981,Bellazzini2017,Ema2017,Calibbi2017,Irastorza2018}. Both axions and ALPs are well-motivated dark-matter candidates \cite{Preskill1983,Abbott1983,Dine1983,Marsh2016}. Their relevant interaction with electromagnetic fields is described by the Lagrangian density
\begin{equation}
\mathcal{L}_{a\gamma\gamma} = -\frac{1}{4}\gagg\,\phi F_{\mu\nu}\tilde{F}^{\mu\nu} = \gagg\,\phi\,\mathbf{E}\cdot\mathbf{B},
\end{equation}
where $\phi$ is the axion field and $\gagg$ is the axion-photon coupling. We use $\hbar=c=1$ in analytical expressions unless SI units are shown explicitly.

In a conventional light-shining-through-a-wall (LSW) experiment, photons convert into axions in a magnetic field before an opaque wall. The axions traverse the wall and convert back into photons in a second magnetic field \cite{Sikivie1983,van_bibber_proposed_1987,Redondo2011}. Optical photon-regeneration experiments and resonantly enhanced designs such as ALPS-II implement this concept \cite{robilliard_no_2007,chou_search_2008,afanasev_experimental_2008,pugnat_results_2008,Ehret2010,ballou_new_2015,Bahre2013}. Interaction between X-ray free-electron laser and crystals extends laboratory searches for axions to higher masses \cite{halliday_bounds_2025}. In vacuum, photons obey $\omega^2=k^2$, whereas axions obey $\omega^2=k^2+m_a^2$. The resultant phase mismatch suppresses coherent conversion once the accumulated phase slip becomes significant, thereby limiting the sensitivity of conventional optical LSW searches for axions at eV-scale masses.

Plasma wakefields provide a complementary environment for axion generation and detection. Laser-driven plasma wakefields contain quasistatic electric fields exceeding $10^{11}\,\mathrm{V/m}$ \cite{Esarey2009,mangles_monoenergetic_2004,geddes_high-quality_2004,faure_laserplasma_2004,blumenfeld_energy_2007}. Plasma channels can also guide intense laser pulses beyond the Rayleigh range and thereby extend the interaction length \cite{miao_optical_2020,picksley_matched_2024,miao_benchmarking_2024,shrock_guided_2024}. These capabilities have motivated studies of axion-plasmon mixing and axion generation in laser-plasma interactions \cite{tercas_axion-plasmon_2018,burton_plasma-based_2018,mendonca_axion_2020,mendonca_axion_2020-1,huang_axion-like_2022,an_situ_2025}. In addition, an underdense plasma changes the photon dispersion to $\omega^2=k^2+\omega_p^2$, where $\omega_p$ is the plasma frequency. Varying the plasma density therefore changes the photon momentum at fixed frequency and can control phase matching between the photon and axion. Axion electrodynamics implemented in particle-in-cell (PIC) simulations provides a direct way to evaluate the resulting generation process \cite{an_modeling_2024}.

Our previous work explored the generation of nearly massless axions using short laser pulses in low-density plasmas \cite{an_situ_2025}. In contrast, the present study focuses on eV-scale axions in a regime of longer laser pulses and higher plasma densities. We develop a complete magnet-free LSW scheme that addresses both the generation and detection of such axions.

\begin{figure}[htbp]
\centering
\includegraphics[width=\linewidth]{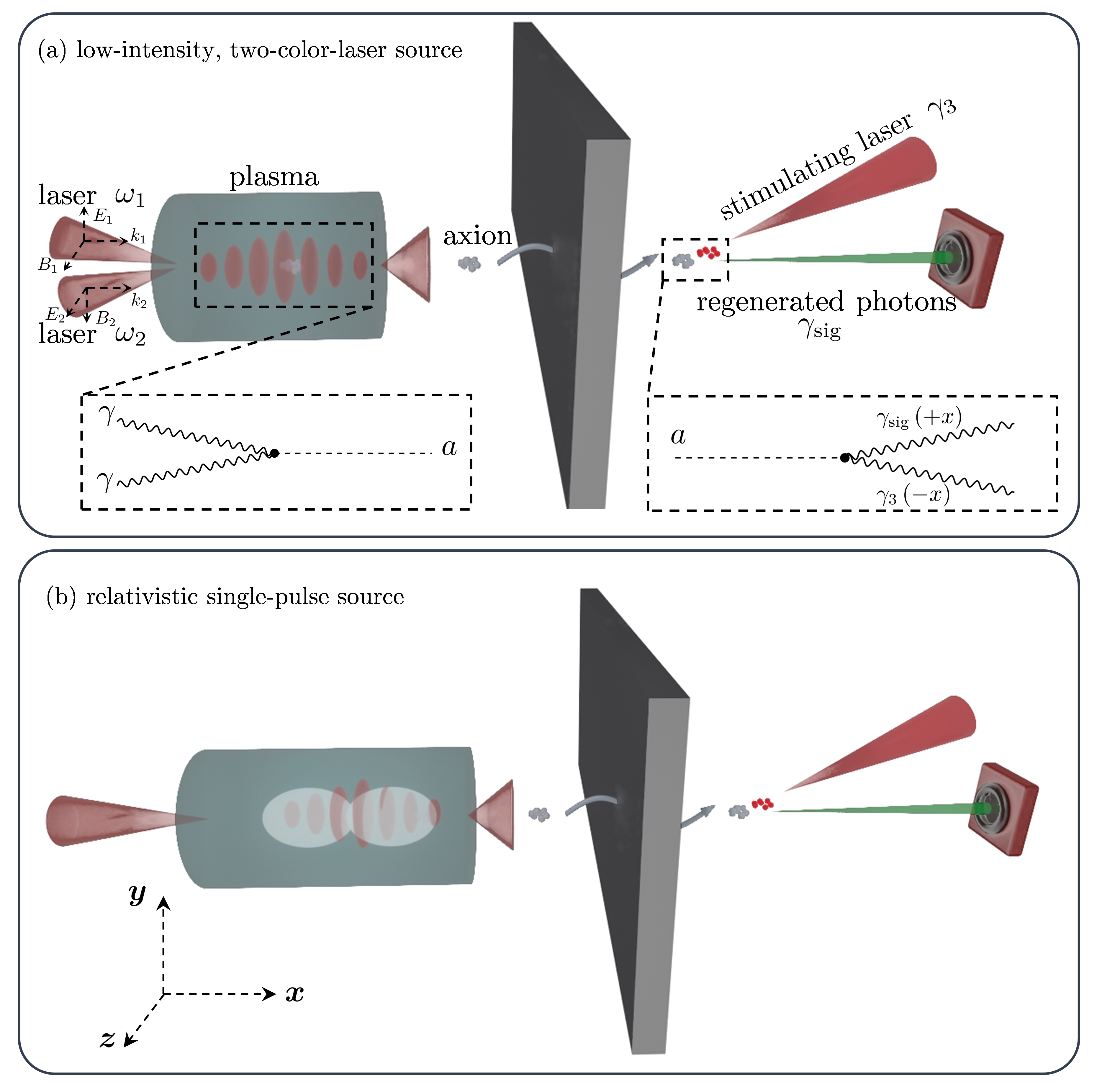}
\caption{Conceptual layouts of the plasma-based LSW scheme using (a) a low-intensity, two-color-laser source and (b) a relativistic single-pulse source. In both configurations, axions generated in an underdense plasma cross the wall and encounter a counter-propagating stimulating laser. The axion decay produces a signal photon propagating in the $+x$ direction and a second photon emitted into the laser mode.}
\label{fig:concept}
\end{figure}

The proposed configurations are illustrated in Fig.~\ref{fig:concept}. Figure~\ref{fig:concept}(a) shows a low-intensity, two-color-laser source with dispersion tailored to a specific axion mass, whereas Fig.~\ref{fig:concept}(b) shows a relativistic single-pulse source whose plasma-induced spectral broadening enables coverage of a broader mass range. In both configurations, the generated axions pass through the wall and encounter a counter-propagating laser that stimulates their decay into a signal photon and a second photon emitted into the laser mode.

\section{Two-Color-Laser Axion Generation}

\subsection{Theoretical analysis for axion generation}

We first consider two laser pulses co-propagating in the $+x$ direction through an underdense plasma. The first has frequency $\omega_1$, wave number $k_1$, and polarization along the $y$ axis. The second has $\omega_2>\omega_1$, wave number $k_2$, and polarization along the $z$ axis. The axion field satisfies
\begin{equation}
\left(\partial_t^2 - \nabla^2 + m_a^2\right)\phi = \gagg\,\mathbf{E}\cdot\mathbf{B}.
\label{eq:kg}
\end{equation}
We define $E_0=m_ec\omega_0/e$ and $B_0=E_0/c$, where $\omega_0$ is the reference laser frequency used for normalization. The normalized electric-field amplitudes of both waves are $a_i=eE_i/(m_ec\omega_0)=E_i/E_0$. Their cross term is
\begin{equation}
\begin{aligned}
\frac{\mathbf{E}\cdot\mathbf{B}}{E_0B_0} ={}& \frac{a_1a_2}{2}
\left(\frac{c}{v_{p,1}}-\frac{c}{v_{p,2}}\right) \\
&\times \left[\cos\bigl((\omega_1+\omega_2)t-(k_1+k_2)x\bigr) \right. \\
&\left. + \cos\bigl((\omega_2-\omega_1)t-(k_2-k_1)x\bigr)\right],
\end{aligned}
\label{eq:source}
\end{equation}
where $v_{p,i}=\omega_i/k_i$. For co-propagating plane waves in vacuum, $v_{p,1}=v_{p,2}=c$ and the source vanishes. However, in the plasma, the plasma dispersion makes the two phase velocities unequal and the source term is nonzero.

\begin{figure}[htbp]
\centering
\includegraphics[width=\linewidth]{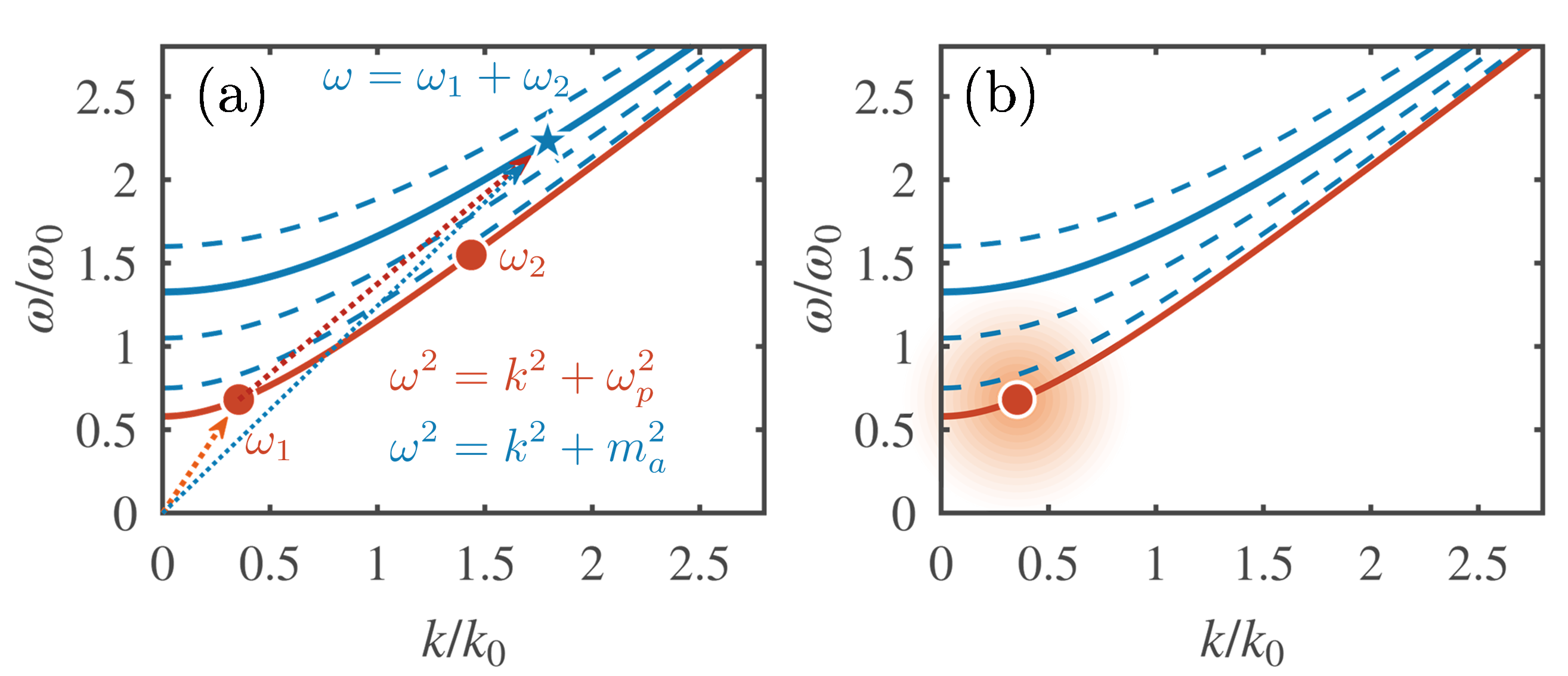}
\caption{Schematic dispersion relations for plasma-based axion generation. The orange solid curve shows the plasma-modified photon dispersion. The blue curves show axion dispersion relations for the matched mass and representative off-resonant masses. (a) Energy--momentum matching for the two-color sum-frequency component in the longitudinal $(k_x,\omega)$ plane. (b) Spectral broadening around a relativistic laser component.}
\label{fig:dispersionconcept}
\end{figure}

We focus on the sum-frequency component. Energy--momentum matching requires this component to lie on the axion dispersion relation
\begin{equation}
m_a^2 = (\omega_1+\omega_2)^2 - (k_1+k_2)^2.
\label{eq:sumbranch}
\end{equation}
For $\omega_i^2=k_i^2+\omega_p^2$ and $\omega_p\ll\omega_i$,
\begin{equation}
m_a^2 \simeq \omega_p^2\,\frac{(\omega_1+\omega_2)^2}{\omega_1\omega_2}.
\label{eq:matchapprox}
\end{equation}
For $\lambda_1=800\,\mathrm{nm}$, $\lambda_2=400\,\mathrm{nm}$, and $n_0=0.01n_c$, where $n_c=m_e\varepsilon_0\omega_0^2/e^2$ is the critical density at $800\,\mathrm{nm}$, Eq.~(\ref{eq:matchapprox}) gives $m_a\simeq0.33\,\mathrm{eV}$. Figure~\ref{fig:dispersionconcept}(a) illustrates the energy--momentum matching in the longitudinal $(k_x,\omega)$ plane.

The matched source generates axions efficiently only near the mass selected by the plasma dispersion. However, at relativistic intensity, the laser drives a nonlinear plasma response that can strongly modulate the pulse \cite{Esarey1994}. This modulation can broaden the Fourier spectrum of $\mathbf{E}\cdot\mathbf{B}$ around the laser branch and provide spectral components that overlap with axion dispersion curves over a wider mass range, as illustrated schematically in Fig.~\ref{fig:dispersionconcept}(b). The PIC simulations below examine axion generation in these two regimes.

\subsection{Simulation results}

We have implemented the axion field and its interaction with the electromagnetic fields in the PIC code EPOCH \cite{arber_contemporary_2015,an_modeling_2024}. In the quasi-cylindrical implementation, the electromagnetic and axion fields are advanced using an X-dispersionless direction-splitting solver \cite{an_xdispersionless_2026}. We use this implementation to solve Eq.~(\ref{eq:kg}) together with the plasma and electromagnetic dynamics.

The simulations retain the first two azimuthal modes after making a Fourier transformation in the azimuthal direction~\cite{an_xdispersionless_2026}. The reference wavelength is $\lambda_0=800\,\mathrm{nm}$, with $k_0=2\pi/\lambda_0=\omega_0/c$ and $T_0=2\pi/\omega_0$. The background electron density is $n_0=0.01n_c$. The transverse laser profile is Gaussian with waist $w_{0}=40\lambda_0$. In the two-color simulations, the second wavelength is $400\,\mathrm{nm}$, the polarizations are orthogonal, and the electric-field amplitudes are equal. Under the common normalization defined above, $a_1=a_2=a_0$. The low-intensity matched source uses $a_0=0.01$. To diagnose relativistic spectral broadening while retaining a direct comparison with the matched source, we first keep the two-color configuration and increase the normalized amplitudes of the two lasers to $a_0=5$. The temporal envelope is
\begin{equation*}
a_i(t)=
\begin{cases}
a_0\sin^2\!\left(\dfrac{\pi t}{10T_0}\right), & 0\leq t<5T_0,\\
a_0, & 5T_0\leq t\leq105T_0,\\
a_0\sin^2\!\left[\dfrac{\pi(110T_0-t)}{10T_0}\right], & 105T_0<t\leq110T_0.
\end{cases}
\end{equation*}
Unless otherwise specified, the final axion energies are evaluated after approximately $2.6\,\mathrm{mm}$ of propagation.

The simulated axion masses are $m_a/\mathrm{eV}=0$, $10^{-3}$, $10^{-2}$, and $0.1$--$1.0$ in steps of 0.1. We set $\gagg=10^{-9}\,\mathrm{GeV}^{-1}$. The axion energy is $\mathcal E_a=\int\mathcal H_a\dd V$, where the energy density is
\begin{equation}
\mathcal H_a=\frac{\hbar}{2c}\left[
\frac{1}{c^2}\left|\partial_t\phi\right|^2
+\left|\nabla\phi\right|^2
+\frac{m_a^2c^2}{\hbar^2}\phi^2\right].
\end{equation}
Because the interaction between the axion and electromagnetic fields is extremely weak, the axion-field amplitude scales as $\gagg$ and the axion energy scales as $\gagg^2$.
We have verified numerically that the simulated axion energy is proportional to $\gagg^2$.

\begin{figure}[htbp]
\centering
\includegraphics[width=\linewidth]{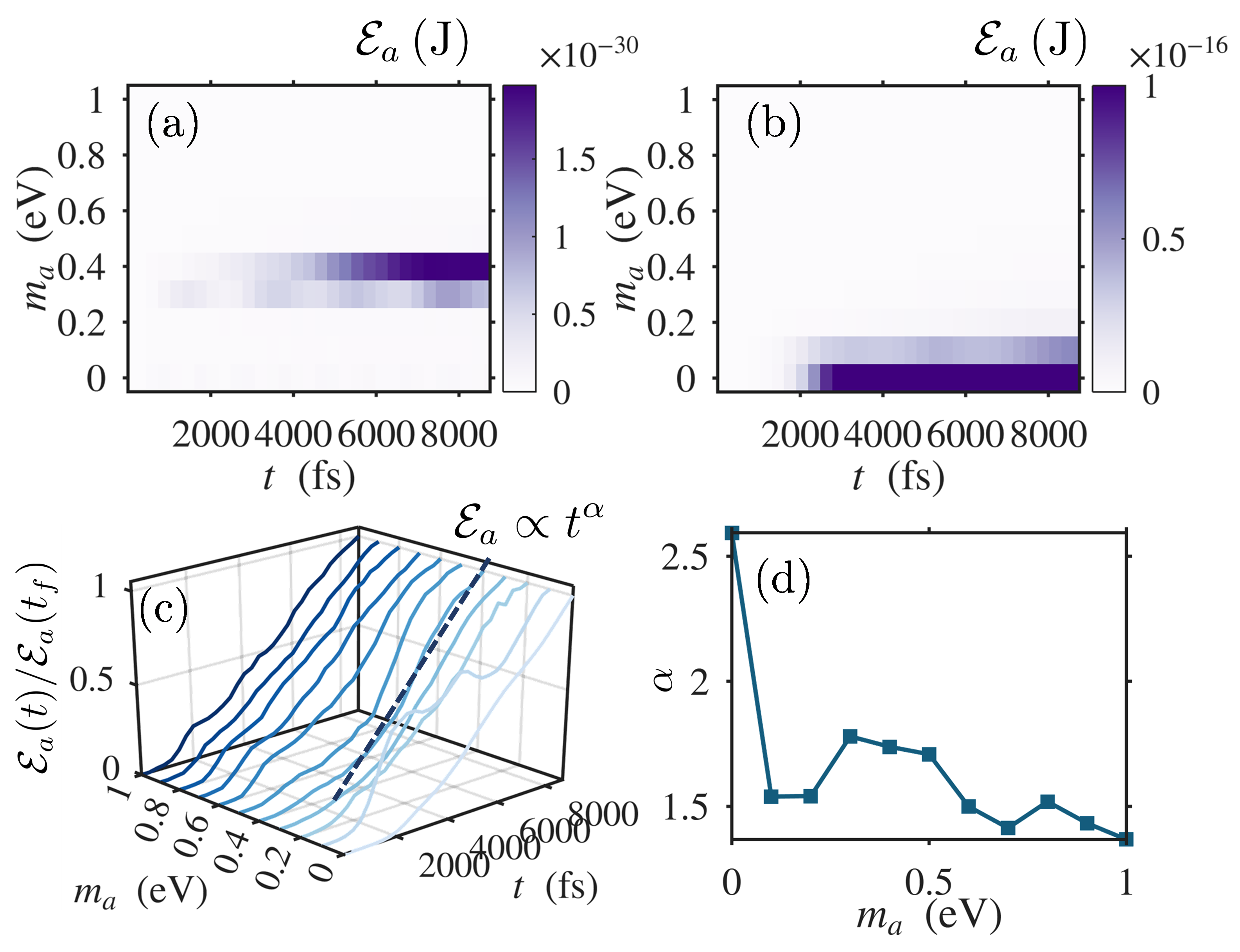}
\caption{Time evolution of the generated axion energy for different masses in the two-color configuration. (a) Matched source with $a_0=0.01$. (b) Relativistic configuration with $a_0=5$. (c) Energy at $a_0=5$ normalized separately for each mass to its final value. (d) Exponent $\alpha$ obtained from the power-law fit $\mathcal E_a\propto t^\alpha$.}
\label{fig:massmaps}
\end{figure}

Figure~\ref{fig:massmaps}(a) shows that generation by the matched source is concentrated near $m_a\simeq0.3\,\mathrm{eV}$, consistent with the value of approximately $0.33\,\mathrm{eV}$ predicted by Eq.~(\ref{eq:matchapprox}). At $a_0=5$, most of the generated axion energy is concentrated at $m_a\lesssim0.1\,\mathrm{eV}$, as shown in Fig.~\ref{fig:massmaps}(b). However, after each curve is normalized to its final energy, Fig.~\ref{fig:massmaps}(c) shows that the axion energy continues to grow throughout the scanned range from 0 to $1\,\mathrm{eV}$. To quantify this growth, we fit the temporal evolution of the axion energy to $\mathcal E_a\propto t^\alpha$.
Figure~\ref{fig:massmaps}(d) shows that $\alpha$ is slightly larger than 1 over the scanned mass range. Axion generation therefore continues during this interval, although the energy grows more slowly than the ideal coherent scaling $\mathcal E_a\propto t^2$. Therefore, the axion energies of different masses can increase over time, and we can expect more axion generation for longer propagation distances.

\begin{figure}[t]
\centering
\includegraphics[width=\linewidth]{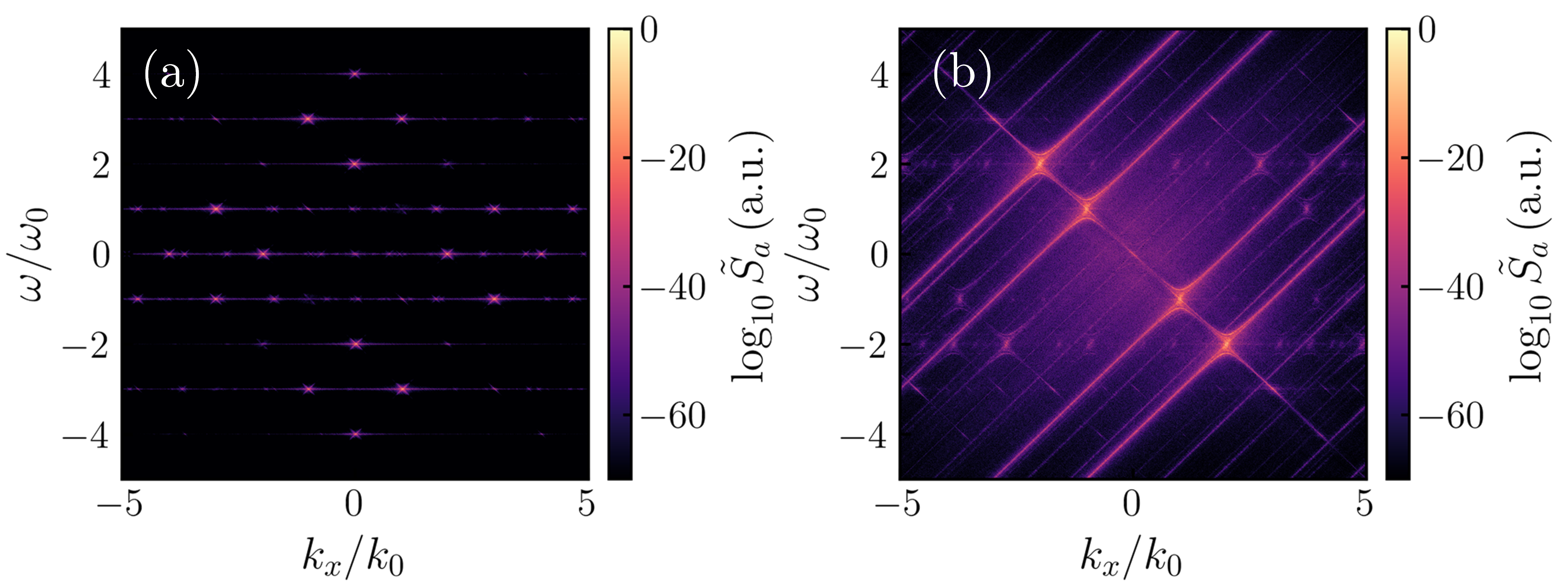}
\caption{Fourier spectra of the axion source term for the two-color configuration at (a) $a_0=0.01$ and (b) $a_0=5$.}
\label{fig:fft}
\end{figure}

To understand the mechanism underlying axion generation, Fourier transformation of the source $S_a\equiv\mathbf E\cdot\mathbf B$ is performed. Figure~\ref{fig:fft}(a) shows a localized sum-frequency component associated with two-color matching at low intensity. However, at relativistic intensity, the spectrum is broadened around the laser branches, as shown in Fig.~\ref{fig:fft}(b). The spectral peaks near the fundamental and second-harmonic laser frequencies indicate that these components arise from mixing between the laser fields and low-frequency plasma waves. Our previous work similarly generated axions through the interaction of the laser electromagnetic field with the electromagnetic fields of a strongly nonlinear plasma wakefield~\cite{an_situ_2025}. In that regime, the laser drove a nearly electron-free plasma bubble and the generation was directed toward nearly massless axions. Here, the longer pulse and higher plasma density sustain stronger modulation of the laser-associated source spectrum. The resulting broadening provides spectral components that overlap with axion dispersion curves over an extended eV-scale mass range and thereby accounts for generation away from the low-intensity resonance.

\begin{figure}[t]
\centering
\includegraphics[width=\linewidth]{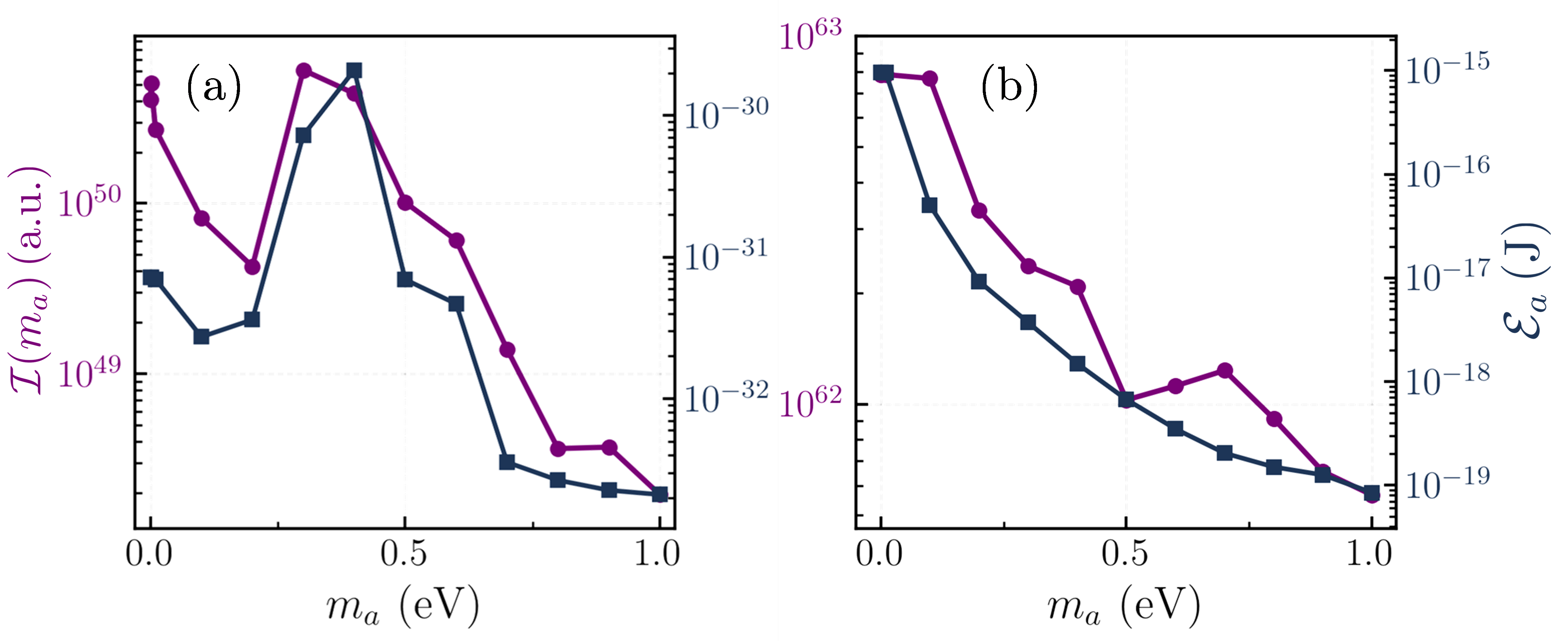}
\caption{Dispersion-overlap diagnostic and final axion energy versus mass for (a) $a_0=0.01$ and (b) $a_0=5$.}
\label{fig:dispersion}
\end{figure}

The two-dimensional Fourier transformation of the axion source term $S_a(t,x) = \mathbf E(t,x)\cdot \mathbf B(t,x)$ is defined as
\begin{equation}
\widetilde S_a(\omega,k)=\int\dd t\dd x\,
S_a(t,x)e^{\mathrm{i}(\omega t-kx)},
\end{equation}
where $k$ is the longitudinal wave number conjugate to $x$. To understand the generation mechanism for such large mass ranges and get the distribution of generated axions, we then define the dispersion-overlap integral and calculate the axion mass spectrum
\begin{equation}
\mathcal I(m_a)=\int_{\omega^2=k^2+m_a^2}
\left|\widetilde S_a(\omega,k)\right|^2\dd k.
\label{eq:dispint}
\end{equation}
At low intensity, both $\mathcal I$ and the final axion energy peak near the matched mass, as shown in Fig.~\ref{fig:dispersion}(a). In the relativistic laser regime, both quantities decrease overall with mass, as shown in Fig.~\ref{fig:dispersion}(b). Their agreement supports the interpretation that the broad mass generation arises from the overlap between the broadened source spectrum and the axion dispersion curves.

\section{Single-pulse axion generation and parameter dependence}
\label{sec:singlepulse}

\begin{figure}[t]
\centering
\includegraphics[width=\linewidth]{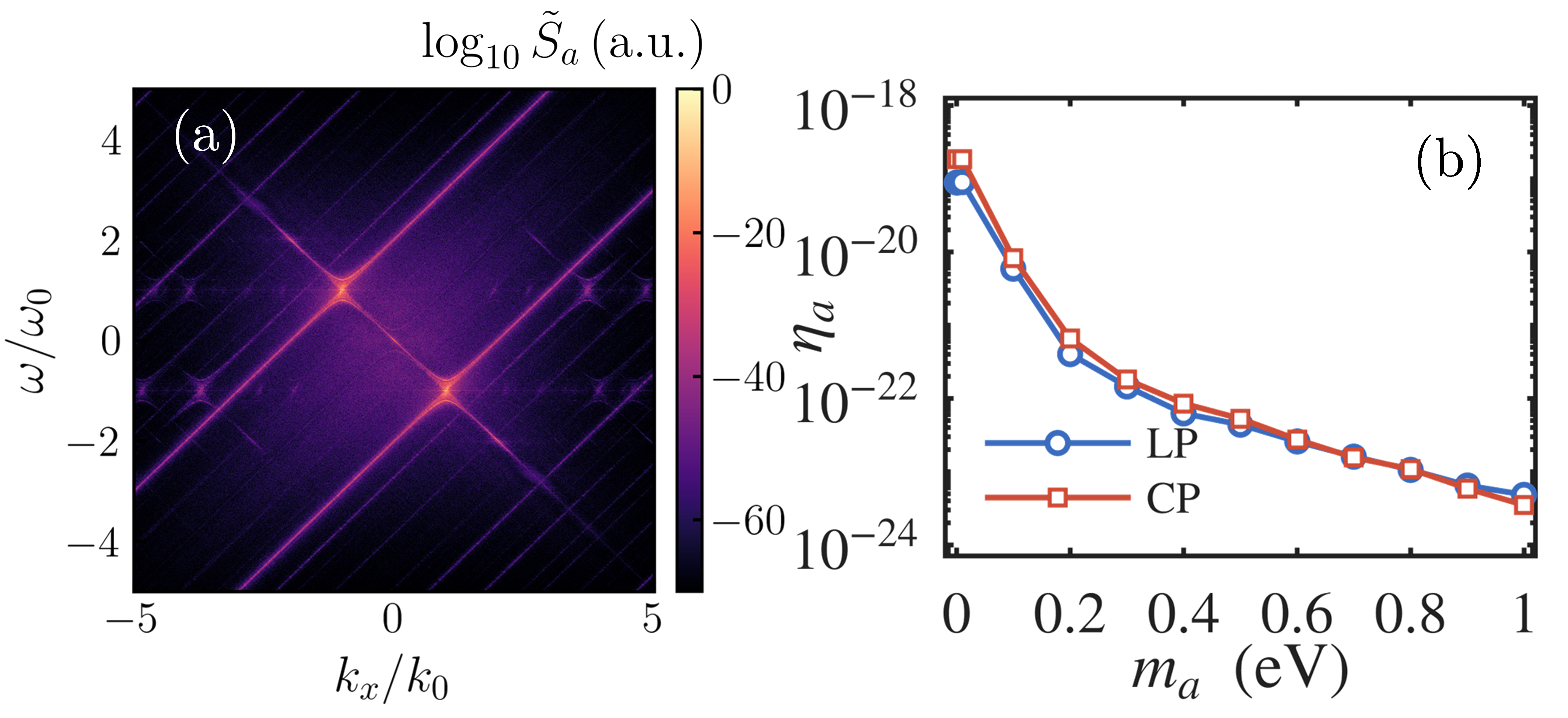}
\caption{Relativistic single-pulse broadband source. (a) Source-term spectrum for a CP pulse. (b) Laser-to-axion conversion efficiency for CP and LP pulses at fixed $a_0$.}
\label{fig:singlepulse}
\end{figure}

The two-color results show that relativistic plasma modulation can broaden the source spectrum beyond the narrow dispersion-matched resonance. This raises the question of whether the second laser color is necessary for broadband axion generation. Figure~\ref{fig:singlepulse}(a) shows that a single circularly polarized $800\,\mathrm{nm}$ pulse at $a_0=5$ produces similar broad structures around the laser branch. As in the relativistic two-color configuration, interaction between the laser and low-frequency plasma waves produces a nonzero axion source, while plasma-induced broadening supplies spectral components that overlap with axion dispersion curves over a broad mass range. The single-pulse configuration therefore realizes the broadband source without a second color and is used for the generation stage in the constraint estimate below.

For single-pulse axion generation, we also consider how the laser polarization affects the generation efficiency. We compare circular polarization (CP) and linear polarization (LP) at fixed $a_0$ using the laser-to-axion conversion efficiency $\eta_{a}=\mathcal E_a/\mathcal E_L$. As shown in Fig.~\ref{fig:singlepulse}(b), the axion conversion efficiency is slightly higher for CP over most of the mass range, while the two polarizations become comparable at the upper end. Thus, we choose CP for the single-pulse generation stage.

\begin{figure}[t]
\centering
\includegraphics[width=\linewidth]{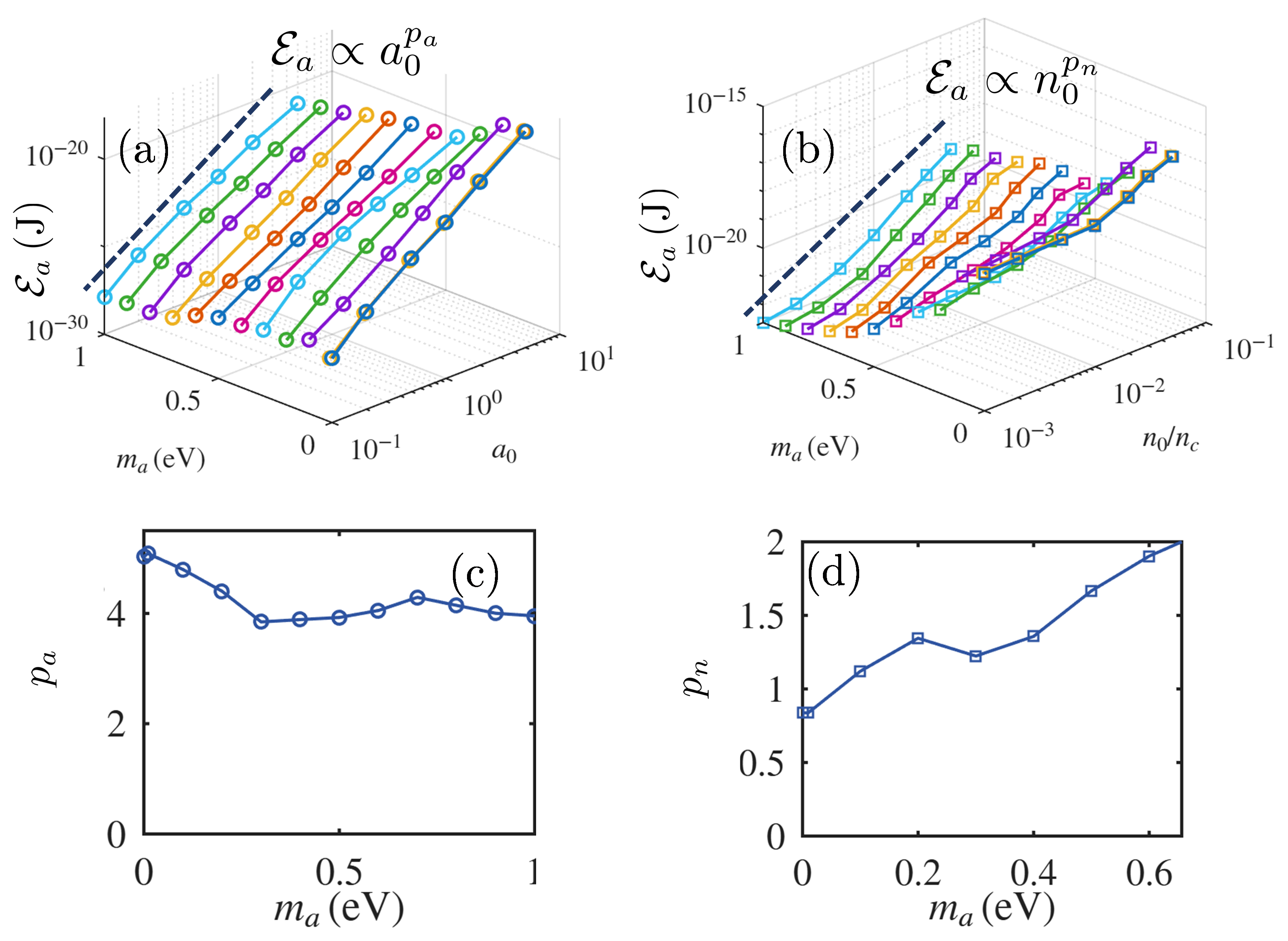}
\caption{Relativistic axion generation as a function of laser amplitude and plasma density after approximately $300\,\mu\mathrm{m}$ of propagation. (a) Final axion energy versus $a_0$ at $n_0=0.01n_c$. (b) Final axion energy versus $n_0$ at $a_0=5$. The total pulse duration is $110T_0$. (c) Effective exponent $p_a$. (d) Effective exponent $p_n$.}
\label{fig:opt_a0_n0}
\end{figure}

Having selected the single-pulse CP configuration, we examine how the laser amplitude and plasma density affect axion generation. Both scans are evaluated after a common propagation distance of approximately $300\,\mu\mathrm{m}$. In Fig.~\ref{fig:opt_a0_n0}(a), $a_0$ varies from 0.1 to 5 at $n_0=0.01n_c$. In Fig.~\ref{fig:opt_a0_n0}(b), $n_0/n_c$ varies from $10^{-3}$ to 0.05 at $a_0=5$. The total pulse duration is $110T_0$ in both scans. For each mass, we obtain the effective exponents $p_a$ and $p_n$ over the scanned intervals from $\mathcal E_a\propto a_0^{p_a}$ and $\mathcal E_a\propto n_0^{p_n}$.

Figure~\ref{fig:opt_a0_n0}(c) shows that $p_a$ varies only weakly with mass and remains mainly between 4 and 5. The density exponent increases overall with mass and reaches approximately 2 at the upper end, as shown in Fig.~\ref{fig:opt_a0_n0}(d). At higher density, the plasma modulates the laser more strongly and produces greater spectral broadening. Heavier axions therefore benefit more from increasing the plasma density.

\begin{figure}[t]
\centering
\includegraphics[width=\linewidth]{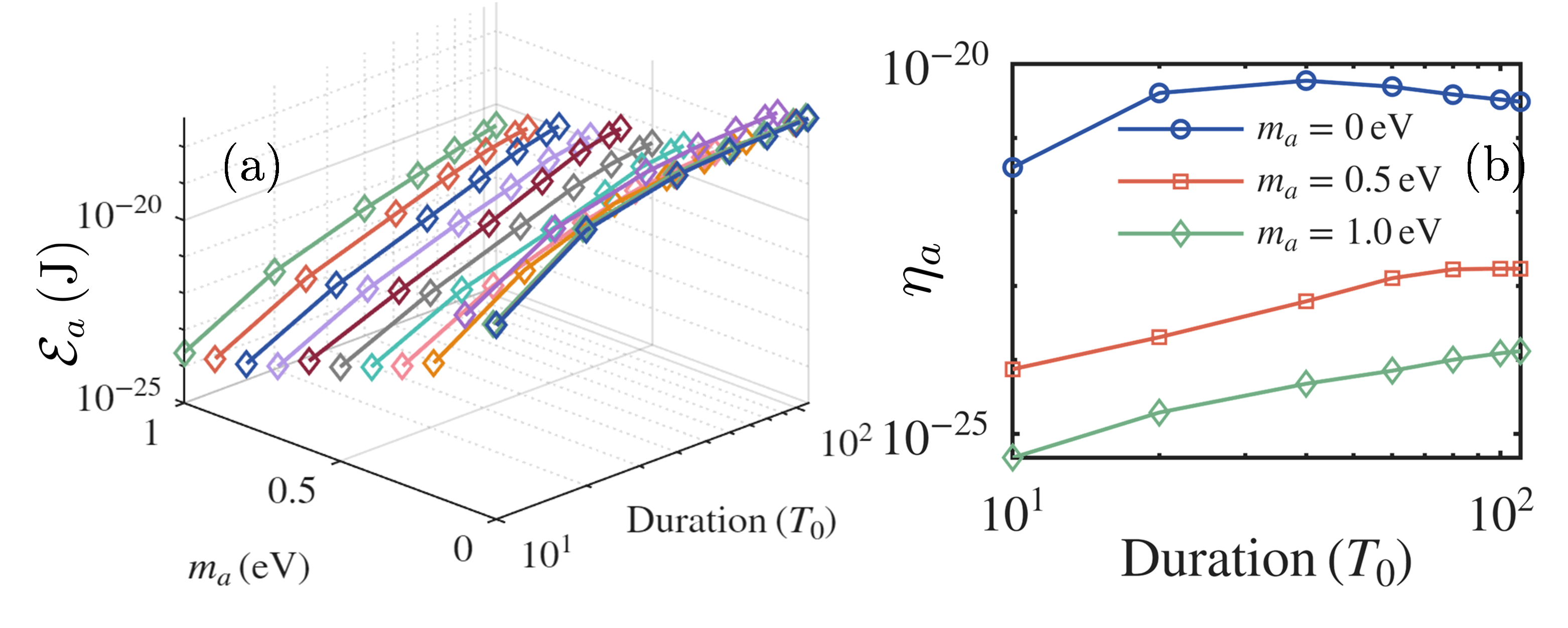}
\caption{Relativistic axion generation as a function of total pulse duration at $a_0=5$ and $n_0=0.01n_c$ after approximately $300\,\mu\mathrm{m}$ of propagation. (a) Generated axion energy for different masses. (b) Laser-to-axion conversion efficiency for $m_a=0$, $0.5\,\mathrm{eV}$, and $1.0\,\mathrm{eV}$.}
\label{fig:opt_duration}
\end{figure}

Besides the laser amplitude and plasma density, we examine the dependence on the total pulse duration from $10T_0$ to $110T_0$ at $a_0=5$ and $n_0=0.01n_c$. Because changing the duration also changes the laser energy, Fig.~\ref{fig:opt_duration}(b) compares the laser-to-axion conversion efficiency. For $m_a=0$, the efficiency rapidly saturates and then decreases slightly. For $m_a=0.5\,\mathrm{eV}$, it rises over most of the scanned interval and approaches saturation at the longest durations. For $m_a=1\,\mathrm{eV}$, it continues to increase throughout the scan. Nearly massless axions do not benefit from plasma-induced spectral broadening, whereas heavier axions are generated by the broadened spectral components. A longer pulse can undergo more plasma modulation and therefore benefits the generation of heavier axions more strongly.

\section{Laser-Stimulated Detection and Projected Constraints}

\subsection{Axion-to-photon conversion}

\begin{figure}
    \centering
    \includegraphics[width=0.4\textwidth]{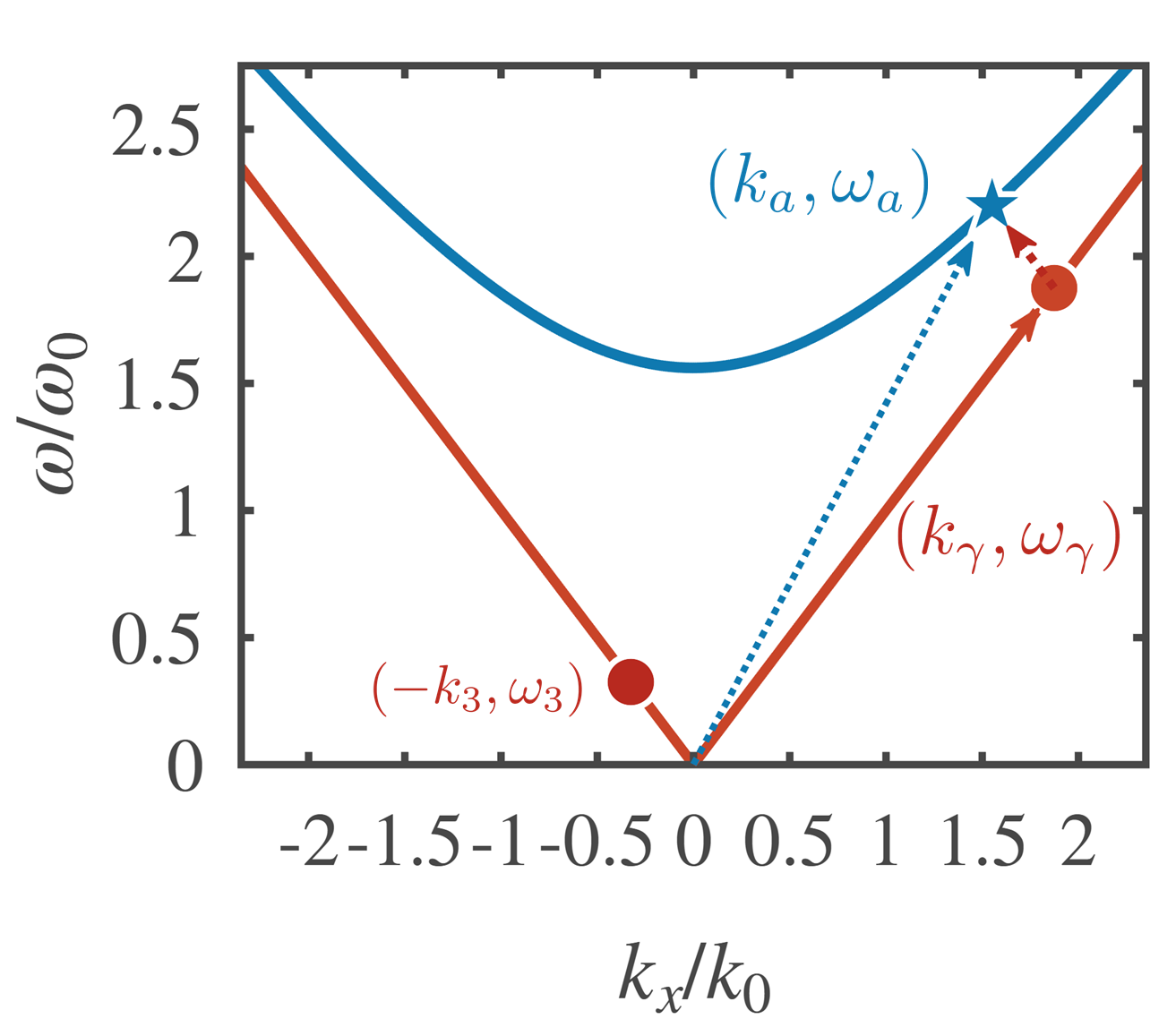}
    \caption{Schematic dispersion relations for the stimulated decay of axions into photons behind the wall. The orange lines show the vacuum photon dispersion, and the blue curve shows the massive-axion dispersion.}
    \label{fig:regeneration_matching}
\end{figure}

To reduce plasma-generated electromagnetic backgrounds and efficiently convert massive axions back into photons, we consider a counter-propagating stimulating laser in vacuum behind the wall. We take $k_3>0$ to denote the magnitude of its wave number, so its longitudinal wave-vector component is $-k_3$. For regenerated photons propagating in the $+x$ direction, energy and longitudinal-momentum conservation give
\begin{equation}
\omega_\gamma = \omega_a - \omega_3, \qquad k_\gamma = k_a + k_3,
\end{equation}
and therefore
\begin{equation}
(\omega_a-\omega_3)^2 = (k_a+k_3)^2.
\label{eq:vacmatch}
\end{equation}
These relations describe stimulated axion decay into a signal photon propagating in the $+x$ direction and a second photon emitted into the counter-propagating stimulating-laser mode. Thus, $\omega_a=\omega_\gamma+\omega_3$, and the signal is at the difference frequency. This is distinct from absorption of a stimulating photon by an axion, which would produce a sum-frequency photon with $\omega'=\omega_a+\omega_3$ and $k'=k_a-k_3$. For the counter-propagating waves in vacuum, this channel gives $\omega'^2-k'^2=m_a^2+2\omega_3(\omega_a+k_a)>0$, so the final photon cannot satisfy its on-shell condition or the dispersion relation in vacuum.
The corresponding energy--momentum matching geometry is illustrated in Fig.~\ref{fig:regeneration_matching}.

With the dispersion relations $\omega_3=k_3$ for the stimulating photon in vacuum and $\omega_a^2=k_a^2+m_a^2$ for the axion, the phase-matched stimulating frequency for a target axion mass at fixed $\omega_a$ is $\omega_3=(\omega_a-k_a)/2$.
At relativistic intensities ($a_0\ge1$), the axions are generated mainly through the interaction of the spectrally broadened laser with low-frequency plasma waves, as shown in Figs.~\ref{fig:fft}(b) and \ref{fig:singlepulse}(a). The axion spectrum is therefore centered on the laser frequency $\omega_0$, with a finite spread. In the constraint estimate below, we adopt the monoenergetic approximation $\omega_a=\omega_0$, corresponding to $\hbar\omega_a=1.55\,\mathrm{eV}$. The axion group velocity is thus mass dependent,
\begin{equation}
\beta_a(m_a)=\frac{k_a}{\omega_a}=\sqrt{1-\frac{m_a^2}{\omega_a^2}}.
\end{equation}
We derive the conversion probability in the small-mixing limit by treating the stimulating laser and axion pulse as plane waves. For the stimulating laser polarization along the $\hat{\mathbf y}$ direction, we write the real fields in terms of their peak complex amplitudes as
\begin{align}
\phi &= \frac{1}{2}\left[\phi_0e^{\mathrm i(k_ax-\omega_at)}+\mathrm{c.c.}\right],\\
\mathbf E_3 &= \frac{1}{2}\left[\hat{\mathbf y}a_3
e^{\mathrm i(-k_3x-\omega_3t)}+\mathrm{c.c.}\right],\\
\mathbf B_3 &= \frac{1}{2}\left[-\hat{\mathbf z}a_3
e^{\mathrm i(-k_3x-\omega_3t)}+\mathrm{c.c.}\right].
\end{align}
In natural units, $a_3$ is the peak amplitude of both the electric and magnetic fields. The complete linearized wave equation for the generated signal field is~\cite{beyer_light-shining-through-wall_2022}
\begin{align}
(\partial_t^2-\nabla^2)\mathbf E_\gamma &= \mathbf S_\gamma,\nonumber\\
\mathbf S_\gamma &\equiv \gagg\partial_t\!\left(
\mathbf B_3\partial_t\phi-\mathbf E_3\times\nabla\phi\right)\nonumber\\
&\quad-\gagg\nabla\!\left[(\nabla\phi)\cdot\mathbf B_3\right].
\label{eq:signalwave}
\end{align}
For the fields above, $(\nabla\phi)\cdot\mathbf B_3=0$. Retaining the difference-frequency component gives
\begin{align}
\mathbf S_\gamma(x,t)
&=\frac{1}{2}\left[\hat{\mathbf z}\widetilde S_\gamma\right.\nonumber\\
&\quad\left.\times e^{\mathrm i(k_\gamma x-\omega_\gamma t)}
+\mathrm{c.c.}\right],\nonumber\\
\widetilde S_\gamma
&=\frac{\gagg}{2}\omega_\gamma(\omega_a+k_a)\phi_0a_3^*.
\label{eq:diffsource}
\end{align}
We introduce the slowly varying signal envelope $E_{\gamma,0}(x)$ through
\begin{equation}
\mathbf E_\gamma(x,t)=\frac{1}{2}\left[\hat{\mathbf z}E_{\gamma,0}(x)
e^{\mathrm i(k_\gamma x-\omega_\gamma t)}+\mathrm{c.c.}\right].
\end{equation}
Using $k_\gamma=\omega_\gamma$ and neglecting the second derivative of the envelope, Eqs.~(\ref{eq:signalwave}) and (\ref{eq:diffsource}) give,
\begin{align}
\frac{\mathrm dE_{\gamma,0}}{\mathrm dx}
&=\frac{\gagg}{4}(\omega_a+k_a)\phi_0a_3^*
e^{\mathrm i\kappa x},\\
E_{\gamma,0}(l_{\text{int}})
&=\frac{\gagg}{4}(\omega_a+k_a)\phi_0a_3^*l_{\text{int}}
e^{\mathrm i\kappa l_{\text{int}}/2}\sinc\!\left(\frac{\kappa l_{\text{int}}}{2}\right),
\label{eq:signalfield}
\end{align}
where $\kappa=k_a+k_3-k_\gamma$ and $l_{\text{int}}$ is the interaction length. The incident axion and output signal-photon energy flux densities are $F_a=\omega_ak_a|\phi_0|^2/2$ and $F_\gamma=|E_{\gamma,0}|^2/2$, respectively. Substituting Eq.~(\ref{eq:signalfield}) and taking the ratio of the number flux densities $F_\gamma/\omega_\gamma$ and $F_a/\omega_a$ yields
\begin{equation}
P_{a\rightarrow\gamma} = \frac{1}{4}\gagg^2 |a_3|^2 l_{\text{int}}^2
\frac{(1+\beta_a)^2}{4\beta_a}\frac{\omega_a}{\omega_\gamma}
\sinc^2\!\left(\frac{\kappa l_{\text{int}}}{2}\right),
\label{eq:pagamma}
\end{equation}
where $\sinc x=\sin x/x$. For $\beta_a\to1$ and $\omega_3\ll\omega_a$, Eq.~(\ref{eq:pagamma}) reduces to $(\gagg |a_3|l_{\text{int}})^2/4$.

We model the stimulating pulse as a focused Gaussian beam with energy $U_L$, waist $w_3$, Rayleigh range $z_R=\pi w_3^2/\lambda_3$, and longitudinal pulse length $l_3=c\tau_d$. Because the stimulating pulse and axions counter-propagate, an axion with velocity $v_a=\beta_a c$ crosses the pulse in a time $l_3/[c(1+\beta_a)]$ and travels only $\beta_a l_3/(1+\beta_a)$ during the overlap. Its finite duration and diffraction therefore jointly limit the effective interaction length to
\begin{equation}
l_{\text{int}}=\min\!\left(\frac{\beta_a}{1+\beta_a}l_3,2z_R\right).
\label{eq:interactionlength}
\end{equation}
For relativistic axions, the pulse-limited interaction length approaches $l_3/2$. In SI units, the magnetic-field amplitude is
\begin{equation}
B_L=2\sqrt{\frac{\mu_0U_L}{\pi w_3^2l_3}}.
\label{eq:laserfield}
\end{equation}
For fixed pulse energy and waist, increasing $l_3$ reduces the magnetic-field amplitude but extends the interaction length until diffraction limits it at $2z_R$. Since the matched probability scales as $B_L^2l_{\text{int}}^2$, it is maximized when the pulse-limited and diffraction-limited lengths are equal. We therefore choose
\begin{equation}
l_{3,{\rm opt}}=2z_R\frac{1+\beta_a}{\beta_a},
\quad \text{then} \quad l_{\text{int}}=2z_R,
\end{equation}
and take $U_L=100\,\mathrm{J}$. Consistent with the generation laser, the base waist is $w_3=32\,\mu\mathrm{m}$. When the stimulating wavelength exceeds this value, the waist is increased to avoid subwavelength focusing.

We use a CO$_2$ laser with $\lambda_3=10.6\,\mu\mathrm{m}$ as the stimulating laser behind the wall. Its pulse length is optimized for $m_a=0.819\,\mathrm{eV}$ and then held fixed for all axion masses. For $w_3=32\,\mu\mathrm{m}$ and $z_R=0.303\,\mathrm{mm}$, this mass corresponds to $\beta_a=0.849$. Equation~(\ref{eq:interactionlength}) gives $l_{\text{int}}=0.607\,\mathrm{mm}$ for an optimal pulse length of $1.322\,\mathrm{mm}$, or a duration of $4.41\,\mathrm{ps}$. For other axion masses, photon regeneration generally remains nonzero and is evaluated by substituting the mass-dependent interaction length from Eq.~(\ref{eq:interactionlength}) into the conversion probability in Eq.~(\ref{eq:pagamma}), while keeping the laser parameters fixed.

\subsection{Coupling constraints}

To obtain the upper limit on the axion-photon coupling $\gagg$ under the assumption that one regenerated photon can be detected, we take the generation stage to be the single-pulse CP configuration with $\lambda_0=800\,\mathrm{nm}$, $a_0=5$, $n_0=0.01n_c$, $w_0=40\lambda_0=32\,\mu\mathrm{m}$, and a total duration of $110T_0\simeq293\,\mathrm{fs}$.

The simulated propagation distance is approximately $2.6\,\mathrm{mm}$. For an assumed $1\,\mathrm{m}$ guided-generation scenario, we linearly extrapolate the simulated axion-energy growth to $1\,\mathrm{m}$. The resulting axion number is
\begin{equation}
N_a^{(1\,\mathrm{m})}(m_a)=\frac{\mathcal E_a^{(1\,\mathrm{m})}(m_a)}{\hbar\omega_0}.
\label{eq:axionnumber}
\end{equation}

The linear extrapolation is motivated by the temporal evolution resolved within the simulated distance. Figure~\ref{fig:massmaps}(c) shows sustained axion-energy growth for all simulated masses, while the fitted exponents in Fig.~\ref{fig:massmaps}(d) remain slightly above 1. The relativistic two-color simulations therefore support approximately linear accumulation over the millimeter-scale interval. Together with the similar broadband source spectrum produced by the single-pulse configuration in Fig.~\ref{fig:singlepulse}(a), this behavior provides the local scaling adopted for the constraint estimate. Extending it to $1\,\mathrm{m}$ assumes that guiding maintains a stable laser envelope and broadened source spectrum, and hence a similar axion-energy yield per unit length. Meter-scale low-density plasma waveguides and high-power guiding over tens of centimeters demonstrate the feasibility of extending the optical interaction well beyond a Rayleigh range \cite{miao_optical_2020,picksley_matched_2024}. Over longer distances, mode mismatch, relativistic self-focusing, pump depletion, self-modulation, and hosing can change the laser spot size, intensity, and spectrum \cite{Esarey2009,Esarey1994}. Evolution of the channel density profile and its ponderomotive modification can also alter the guided modes and produce sustained mode beating \cite{miao_benchmarking_2024,shrock_guided_2024}. These processes can enhance source broadening in some regions and reduce the usable intensity in others, leading to departures from linear axion-energy growth. Fully determining their cumulative effect requires a complete $1\,\mathrm{m}$ propagation simulation with a realistic plasma channel profile. The present millimeter-scale simulations nevertheless self-consistently resolve the nonlinear laser--plasma dynamics, the associated source broadening, and sustained axion generation. They therefore provide a physically grounded quantitative basis and local growth law for the conditional meter-scale projection.

We next specify the configuration used to obtain the coupling constraint. A 30-day run at $1\,\mathrm{Hz}$ contains $N_{\rm shot}=2.592\times10^6$ shots, all using the same CO$_2$ laser setting. The expected regenerated-photon number at the reference coupling is
\begin{equation}
N_\gamma^{\mathrm{reg}}(m_a) = N_{\rm shot}\,N_a^{(1\,\mathrm{m})}(m_a)\,P_{a\rightarrow\gamma}(m_a),
\label{eq:yield}
\end{equation}
Axion generation and axion-to-photon conversion each scale as $\gagg^2$, so the expected regenerated-photon number scales as $\gagg^4$. Assuming that one regenerated photon is detectable, the projected upper limit on the coupling is obtained by requiring
\begin{equation}
N_\gamma^{\mathrm{reg}}\!\left(m_a\right)\ge 1.
\label{eq:glim}
\end{equation}

\begin{figure}[t]
\centering
\includegraphics[width=\linewidth]{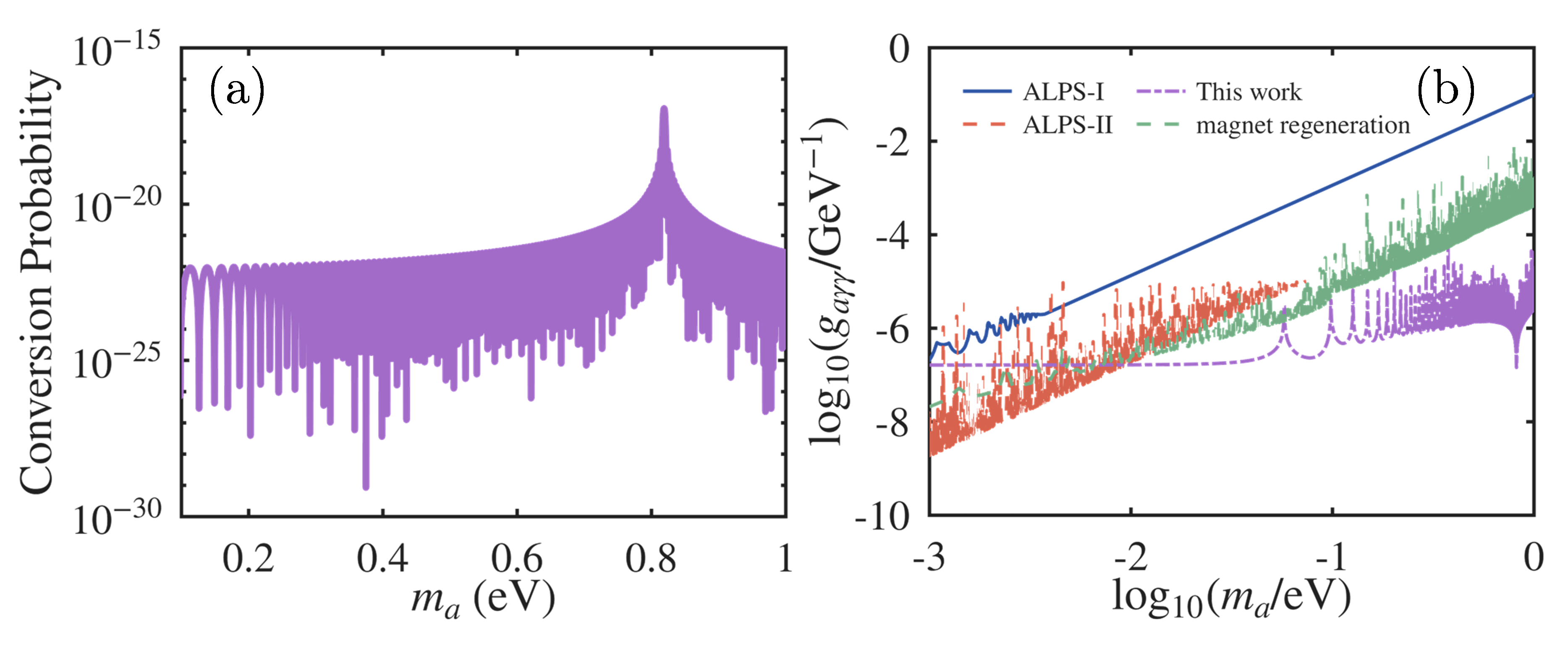}
\caption{Regeneration schemes and projected coupling constraints. (a) Axion-to-photon conversion probability for the CO$_2$ laser. (b) Projected upper limit on $\gagg$, together with the ALPS-I bound and ALPS-II projection from Ref.~\cite{AxionLimits}. The magnet-regeneration curve uses a $1\,\mathrm{T}$ static field over $10\,\mathrm{m}$ behind the wall, while the plasma-generated axion source is unchanged.}
\label{fig:regeneration}
\end{figure}

Figure~\ref{fig:regeneration}(a) shows the axion-to-photon conversion probability for the CO$_2$ laser. It peaks at the matched mass $m_a=0.819\,\mathrm{eV}$ and oscillates away from the peak because of the $\sinc^2$ phase-matching factor. Figure~\ref{fig:regeneration}(b) compares the projected limits for laser-stimulated and magnet regeneration. The static-field curve is strongly suppressed as phase mismatch accumulates along the magnet. The stimulating laser is phase matched at the target mass, and its shorter interaction length limits phase slip away from exact matching. Thus, the eV-scale reach in this work relies on both broadband axion generation before the wall and effective regeneration behind it. At the matched mass, the projected coupling constraint is $g_{a\gamma\gamma}^{\mathrm{lim}}=1.44\times10^{-7}\,\mathrm{GeV}^{-1}$.

Combining Eqs.~(\ref{eq:yield}), (\ref{eq:pagamma}), (\ref{eq:interactionlength}), and (\ref{eq:laserfield}), the dependence on the experimental parameters can be written in a single scaling relation,
\begin{align}
g_{a\gamma\gamma}^{\mathrm{lim}}
\propto \Bigg[& RT_{\rm run}
\frac{\mathcal E_a(a_0,n_0,\tau,L_{\rm g})}{\gagg^2}
\frac{U_L}{w_3^2l_3}
 l_{\mathrm{int}}^2
\frac{(1+\beta_a)^2}{4\beta_a}
\nonumber\\[-2pt]
&\times\sinc^2\!\left(
\frac{\kappa l_{\mathrm{int}}}{2}
\right)\Bigg]^{-1/4}.
\label{eq:glim_regeneration_scaling}
\end{align}
Here $R$ is the repetition rate and $\mathcal E_a/\gagg^2$ is independent of the coupling. Under the linear extrapolation adopted here, $\mathcal E_a\propto L_{\rm g}$ for the guided-generation length $L_{\rm g}$, which gives $g_{a\gamma\gamma}^{\mathrm{lim}}\propto L_{\rm g}^{-1/4}$. The parameter scans in Fig.~\ref{fig:opt_a0_n0} further give $g_{a\gamma\gamma}^{\mathrm{lim}}\propto a_0^{-p_a/4}n_0^{-p_n/4}$ over the simulated intervals. With $p_a\simeq4$--$5$, this corresponds to approximately $g_{a\gamma\gamma}^{\mathrm{lim}}\propto a_0^{-1}$--$a_0^{-5/4}$. For heavier axions, for which $p_n$ approaches 2, the density dependence approaches $g_{a\gamma\gamma}^{\mathrm{lim}}\propto n_0^{-1/2}$. In the generation stage, increasing $\tau$ at fixed $a_0$ and waist first increases the laser energy in proportion to $\tau$, giving the baseline scaling $g_{a\gamma\gamma}^{\mathrm{lim}}\propto\tau^{-1/4}$ if the laser-to-axion conversion efficiency is unchanged. Figure~\ref{fig:opt_duration}(b) shows that this efficiency also increases with duration for heavier axions. Their constraint therefore improves more strongly than $\tau^{-1/4}$ over the simulated duration range.

In Eq.~(\ref{eq:glim_regeneration_scaling}), $w_3$ and $l_3$ determine both the stimulating field amplitude and the interaction length. Near phase matching, the pulse-limited regime, $\beta_a l_3/(1+\beta_a)<2z_R$, gives $g_{a\gamma\gamma}^{\mathrm{lim}}\propto U_L^{-1/4}w_3^{1/2}l_3^{-1/4}$. In the diffraction-limited regime, $\beta_a l_3/(1+\beta_a)>2z_R$, the corresponding scaling is $g_{a\gamma\gamma}^{\mathrm{lim}}\propto U_L^{-1/4}w_3^{-1/2}l_3^{1/4}$ at fixed stimulating wavelength. The two regimes meet at $l_{3,{\rm opt}}=2z_R(1+\beta_a)/\beta_a$, which maximizes the matched conversion probability. With the stimulating pulse fixed at the value optimized for $m_a=0.819\,\mathrm{eV}$, the sensitivity at other masses follows from the full phase-matching factor and the mass-dependent interaction length and does not obey a single power law. These scalings show that increasing the generation-laser amplitude and the plasma density provides a stronger improvement in the projected constraint than increasing the running time or stimulating-laser energy, which improve the constraint only through a quarter-power dependence.

\section{Conclusions}

We have studied a magnet-free LSW scheme for the generation and detection of eV-scale axions. For the generation, two configurations have been considered, one based upon two-color-laser interactions and another based upon a single relativistic high power laser pulse. For the two-color-laser configuration, at low intensity, plasma dispersion phase matches a two-color source to a narrow axion mass range. At relativistic intensity, plasma modulation broadens the source spectrum around the laser branch. For the single relativistic laser configuration, axions are generated over a wider mass range. A scan of laser and plasma parameters illustrates how the broadband source can be optimized for heavier axions. Behind the wall, a counter-propagating laser stimulates axion decay into a detectable signal photon and a second photon emitted into the laser mode. Under representative assumptions, requiring at least one regenerated photon yields projected upper limits on $\gagg$ of order $10^{-7}\,\mathrm{GeV}^{-1}$ over $m_a=0.1$--$1\,\mathrm{eV}$.

\begin{acknowledgments}
This work was supported by the National Natural Science Foundation of China (Grants No. 12225505, 12090060, 12090061, and 12605402), the Fundamental and Interdisciplinary Disciplines Breakthrough Plan of the Ministry of Education of China (JYB2025XDXM204), the Office of Science and Technology of the Shanghai Municipal Government (Grant No. 23JC1410200), the Shanghai Jiao Tong University 2030 Initiative, and the State Key Laboratory of Dark Matter Physics. The computations were performed on the $\pi$ 2.0 cluster at the Center for High Performance Computing, Shanghai Jiao Tong University.
\end{acknowledgments}

\bibliographystyle{apsrev4-2}
\bibliography{massive_axion_plasma}

@article{PecceiQuinn1977,
  author = {Peccei, R. D. and Quinn, Helen R.},
  title = {CP Conservation in the Presence of Instantons},
  journal = {Phys. Rev. Lett.},
  volume = {38},
  pages = {1440--1443},
  year = {1977},
  doi = {10.1103/PhysRevLett.38.1440}
}

@article{Weinberg1978,
  author = {Weinberg, Steven},
  title = {A New Light Boson?},
  journal = {Phys. Rev. Lett.},
  volume = {40},
  pages = {223--226},
  year = {1978},
  doi = {10.1103/PhysRevLett.40.223}
}

@article{Wilczek1978,
  author = {Wilczek, Frank},
  title = {Problem of Strong $P$ and $T$ Invariance in the Presence of Instantons},
  journal = {Phys. Rev. Lett.},
  volume = {40},
  pages = {279--282},
  year = {1978},
  doi = {10.1103/PhysRevLett.40.279}
}

@article{PecceiQuinn1977b,
  author = {Peccei, R. D. and Quinn, Helen R.},
  title = {Constraints Imposed by {CP} Conservation in the Presence of Pseudoparticles},
  journal = {Phys. Rev. D},
  volume = {16},
  pages = {1791--1797},
  year = {1977},
  doi = {10.1103/PhysRevD.16.1791}
}

@article{Kim2010,
  author = {Kim, Jihn E. and Carosi, Gianpaolo},
  title = {Axions and the Strong {CP} Problem},
  journal = {Rev. Mod. Phys.},
  volume = {82},
  pages = {557--601},
  year = {2010},
  doi = {10.1103/RevModPhys.82.557}
}

@article{Irastorza2018,
  author = {Irastorza, Igor G. and Redondo, Javier},
  title = {New Experimental Approaches in the Search for Axion-Like Particles},
  journal = {Prog. Part. Nucl. Phys.},
  volume = {102},
  pages = {89--159},
  year = {2018},
  doi = {10.1016/j.ppnp.2018.05.003}
}

@article{Svrcek2006,
  author = {Svrcek, Peter and Witten, Edward},
  title = {Axions in String Theory},
  journal = {J. High Energy Phys.},
  volume = {2006},
  number = {06},
  pages = {051},
  year = {2006},
  doi = {10.1088/1126-6708/2006/06/051}
}

@article{Gelmini1981,
  author = {Gelmini, G. B. and Roncadelli, M.},
  title = {Left-Handed Neutrino Mass Scale and Spontaneously Broken Lepton Number},
  journal = {Phys. Lett. B},
  volume = {99},
  pages = {411--415},
  year = {1981},
  doi = {10.1016/0370-2693(81)90559-1}
}

@article{Bellazzini2017,
  author = {Bellazzini, Brando and Mariotti, Alberto and Redigolo, Diego and Sala, Filippo and Serra, Javi},
  title = {{$R$-Axion at Colliders}},
  journal = {Phys. Rev. Lett.},
  volume = {119},
  number = {14},
  pages = {141804},
  year = {2017},
  doi = {10.1103/PhysRevLett.119.141804}
}

@article{Ema2017,
  author = {Ema, Yohei and Hamaguchi, Koichi and Moroi, Takeo and Nakayama, Kazunori},
  title = {Flaxion: A Minimal Extension to Solve Puzzles in the Standard Model},
  journal = {J. High Energy Phys.},
  volume = {2017},
  number = {01},
  pages = {096},
  year = {2017},
  doi = {10.1007/JHEP01(2017)096}
}

@article{Calibbi2017,
  author = {Calibbi, Lorenzo and Goertz, Florian and Redigolo, Diego and Ziegler, Robert and Zupan, Jure},
  title = {Minimal Axion Model from Flavor},
  journal = {Phys. Rev. D},
  volume = {95},
  number = {9},
  pages = {095009},
  year = {2017},
  doi = {10.1103/PhysRevD.95.095009}
}

@article{Sikivie1983,
  author = {Sikivie, Pierre},
  title = {Experimental Tests of the ``Invisible'' Axion},
  journal = {Phys. Rev. Lett.},
  volume = {51},
  pages = {1415--1417},
  year = {1983},
  doi = {10.1103/PhysRevLett.51.1415}
}

@article{Redondo2011,
  author = {Redondo, Javier and Ringwald, Andreas},
  title = {Light shining through walls},
  journal = {Contemp. Phys.},
  volume = {52},
  number = {3},
  pages = {211--236},
  year = {2011},
  doi = {10.1080/00107514.2011.563516}
}

@article{Preskill1983,
  author = {Preskill, John and Wise, Mark B. and Wilczek, Frank},
  title = {Cosmology of the Invisible Axion},
  journal = {Phys. Lett. B},
  volume = {120},
  pages = {127--132},
  year = {1983},
  doi = {10.1016/0370-2693(83)90637-8}
}

@article{Abbott1983,
  author = {Abbott, L. F. and Sikivie, P.},
  title = {A Cosmological Bound on the Invisible Axion},
  journal = {Phys. Lett. B},
  volume = {120},
  pages = {133--136},
  year = {1983},
  doi = {10.1016/0370-2693(83)90638-X}
}

@article{Dine1983,
  author = {Dine, Michael and Fischler, Willy},
  title = {The Not So Harmless Axion},
  journal = {Phys. Lett. B},
  volume = {120},
  pages = {137--141},
  year = {1983},
  doi = {10.1016/0370-2693(83)90639-1}
}

@article{Marsh2016,
  author = {Marsh, David J. E.},
  title = {Axion Cosmology},
  journal = {Phys. Rept.},
  volume = {643},
  pages = {1--79},
  year = {2016},
  doi = {10.1016/j.physrep.2016.06.005}
}

@article{Esarey2009,
  author = {Esarey, Eric and Schroeder, Carl B. and Leemans, Wim P.},
  title = {Physics of laser-driven plasma-based electron accelerators},
  journal = {Rev. Mod. Phys.},
  volume = {81},
  pages = {1229--1285},
  year = {2009},
  doi = {10.1103/RevModPhys.81.1229}
}

@article{Esarey1994,
  author = {Esarey, Eric and Krall, Jonathan and Sprangle, Phillip},
  title = {Envelope Analysis of Intense Laser Pulse Self-Modulation in Plasmas},
  journal = {Phys. Rev. Lett.},
  volume = {72},
  number = {18},
  pages = {2887--2890},
  year = {1994},
  doi = {10.1103/PhysRevLett.72.2887}
}

@article{Ehret2010,
  author = {Ehret, Klaus and Frede, Martin and Ghazaryan, S. and Hallal, A. and Knabbe, Ernst-Helmut and Kracht, Dieter and Lindner, Axel and Willke, Benno and Zimer, Peter},
  title = {New {ALPS} Results on Hidden-Sector Lightweights},
  journal = {Phys. Lett. B},
  volume = {689},
  pages = {149--155},
  year = {2010},
  doi = {10.1016/j.physletb.2010.04.066}
}

@article{Bahre2013,
  author = {B"ahre, R. and D"obrich, B. and Dreyling-Eschweiler, J. and Gies, H. and Gianotti, F. and others},
  title = {Any light particle search {II} --- Technical Design Report},
  journal = {JINST},
  volume = {8},
  number = {09},
  pages = {T09001},
  year = {2013},
  doi = {10.1088/1748-0221/8/09/T09001}
}

@article{van_bibber_proposed_1987,
  author = {Van Bibber, K. and Dagdeviren, N. R. and Koonin, S. E. and Kerman, A. K. and Nelson, H. N.},
  title = {Proposed experiment to produce and detect light pseudoscalars},
  journal = {Phys. Rev. Lett.},
  volume = {59},
  pages = {759--762},
  year = {1987},
  doi = {10.1103/PhysRevLett.59.759}
}

@article{afanasev_experimental_2008,
  author = {Afanasev, A. and Baker, O. K. and Beard, K. B. and Biallas, G. and Boyce, J. and Minarni, M. and Ramdon, R. and Shinn, M. and Slocum, P.},
  title = {Experimental Limit on Optical-Photon Coupling to Light Neutral Scalar Bosons},
  journal = {Phys. Rev. Lett.},
  volume = {101},
  pages = {120401},
  year = {2008},
  doi = {10.1103/PhysRevLett.101.120401}
}

@article{robilliard_no_2007,
  author = {Robilliard, C. and Battesti, R. and Fouche, M. and Mauchain, J. and Sautivet, A.-M. and Amiranoff, F. and Rizzo, C.},
  title = {No ``Light Shining through a Wall'': Results from a Photoregeneration Experiment},
  journal = {Phys. Rev. Lett.},
  volume = {99},
  pages = {190403},
  year = {2007},
  doi = {10.1103/PhysRevLett.99.190403}
}

@article{chou_search_2008,
  author = {Chou, A. S. and Wester, W. and Baumbaugh, A. and Gustafson, H. R. and Irizarry-Valle, Y. and Mazur, P. O. and Steffen, J. H. and Tomlin, R. and Yang, X. and Yoo, J.},
  title = {Search for Axionlike Particles Using a Variable-Baseline Photon-Regeneration Technique},
  journal = {Phys. Rev. Lett.},
  volume = {100},
  pages = {080402},
  year = {2008},
  doi = {10.1103/PhysRevLett.100.080402}
}

@article{pugnat_results_2008,
  author = {Pugnat, Pierre and Duvillaret, Lionel and Jost, Remy and Vitrant, Guy and Romanini, Daniele and Siemko, Andrzej and Ballou, Rafik and Barbara, Bernard and Finger, Michael and Finger, Miroslav and Hosek, Jan and Kral, Miroslav and Meissner, Krzysztof A. and Sulc, Miroslav and Zicha, Josef},
  title = {Results from the {OSQAR} Photon-Regeneration Experiment: No Light Shining through a Wall},
  journal = {Phys. Rev. D},
  volume = {78},
  pages = {092003},
  year = {2008},
  doi = {10.1103/PhysRevD.78.092003}
}

@article{halliday_bounds_2025,
  author = {Halliday, Jack W. D. and Marocco, Giacomo and Beyer, Konstantin A. and Heaton, Charles and Nakatsutsumi, Motoaki and Preston, Thomas R. and Arrowsmith, Charles D. and Baehtz, Carsten and Goede, Sebastian and Humphries, Oliver and Garcia, Alejandro Laso and Plackett, Richard and Svensson, Pontus and Vacalis, Georgios and Wark, Justin and Wood, Daniel and Zastrau, Ulf and Bingham, Robert and Shipsey, Ian and Sarkar, Subir and Gregori, Gianluca},
  title = {Bounds on Heavy Axions with an {X}-Ray Free Electron Laser},
  journal = {Phys. Rev. Lett.},
  volume = {134},
  pages = {055001},
  year = {2025},
  doi = {10.1103/PhysRevLett.134.055001}
}

@article{ballou_new_2015,
  author = {Ballou, R. and Deferne, G. and Finger, M. and Finger, M. and Flekova, L. and Hosek, J. and Kunc, S. and Macuchova, K. and Meissner, K. A. and Pugnat, P. and Schott, M. and Siemko, A. and Slunecka, M. and Sulc, M. and Weinsheimer, C. and Zicha, J. and OSQAR Collaboration},
  title = {New exclusion limits on scalar and pseudoscalar axionlike particles from light shining through a wall},
  journal = {Phys. Rev. D},
  volume = {92},
  pages = {092002},
  year = {2015},
  doi = {10.1103/PhysRevD.92.092002}
}

@article{burton_plasma-based_2018,
  author = {Burton, David A. and Noble, Adam},
  title = {Plasma-based wakefield accelerators as sources of axion-like particles},
  journal = {New J. Phys.},
  volume = {20},
  pages = {033022},
  year = {2018},
  doi = {10.1088/1367-2630/aab475}
}

@article{huang_axion-like_2022,
  author = {Huang, Shan and Shen, Baifei and Bu, Zhigang and Zhang, Xiaomei and Ji, Liangliang and Zhai, Shuhua},
  title = {Axion-Like Particle Generation in Laser-Plasma Interaction},
  journal = {Phys. Scr.},
  volume = {97},
  pages = {105303},
  year = {2022},
  doi = {10.1088/1402-4896/ac8b6b}
}

@article{tercas_axion-plasmon_2018,
  author = {Tercas, H. and Rodrigues, J. D. and Mendonca, J. T.},
  title = {Axion-Plasmon Polaritons in Strongly Magnetized Plasmas},
  journal = {Phys. Rev. Lett.},
  volume = {120},
  pages = {181803},
  year = {2018},
  doi = {10.1103/PhysRevLett.120.181803}
}

@article{mendonca_axion_2020,
  author = {Mendonca, J. T. and Tercas, H. and Rodrigues, J. D.},
  title = {Axion excitation by intense laser fields in a plasma},
  journal = {Phys. Scr.},
  volume = {95},
  pages = {045601},
  year = {2020},
  doi = {10.1088/1402-4896/ab5ab5}
}

@article{mendonca_axion_2020-1,
  author = {Mendonca, J. T. and Rodrigues, J. D. and Tercas, H.},
  title = {Axion production in unstable magnetized plasmas},
  journal = {Phys. Rev. D},
  volume = {101},
  pages = {051701},
  year = {2020},
  doi = {10.1103/PhysRevD.101.051701}
}

@article{an_modeling_2024,
  author = {An, Xiangyan and Chen, Min and Liu, Jianglai and Sheng, Zhengming and Zhang, Jie},
  title = {Modeling of axion and electromagnetic fields interaction in particle-in-cell simulations},
  journal = {Matter Radiat. Extrem.},
  volume = {9},
  pages = {067204},
  year = {2024},
  doi = {10.1063/5.0226159}
}

@misc{an_xdispersionless_2026,
  author = {An, Xiangyan and Chen, Min and Wu, Yipeng and Liu, Jianglai and Sheng, Zhengming and Zhang, Jie},
  title = {{X}-Dispersionless Solver for Electromagnetic and Axion Fields in a Cylindrical Particle-in-Cell Code},
  year = {2026},
  note = {Submitted to J. Comput. Phys.},
  doi = {10.2139/ssrn.6352078},
  howpublished = {Available at SSRN: \url{https://ssrn.com/abstract=6352078}}
}

@article{arber_contemporary_2015,
  author = {Arber, T. D. and Bennett, K. and Brady, C. S. and Lawrence-Douglas, A. and Ramsay, M. G. and Sircombe, N. J. and Gillies, P. and Evans, R. G. and Schmitz, H. and Bell, A. R. and Ridgers, C. P.},
  title = {Contemporary Particle-in-Cell Approach to Laser-Plasma Modelling},
  journal = {Plasma Phys. Control. Fusion},
  volume = {57},
  pages = {113001},
  year = {2015},
  doi = {10.1088/0741-3335/57/11/113001}
}

@article{mangles_monoenergetic_2004,
  author = {Mangles, S. P. D. and Murphy, C. D. and Najmudin, Z. and Thomas, A. G. R. and Collier, J. L. and Dangor, A. E. and Divall, E. J. and Foster, P. S. and Gallacher, J. G. and Hooker, C. J. and Jaroszynski, D. A. and Langley, A. J. and Mori, W. B. and Norreys, P. A. and Tsung, F. S. and Viskup, R. and Walton, B. R. and Krushelnick, K.},
  title = {Monoenergetic Beams of Relativistic Electrons from Intense Laser-Plasma Interactions},
  journal = {Nature},
  volume = {431},
  pages = {535--538},
  year = {2004},
  doi = {10.1038/nature02939}
}

@article{geddes_high-quality_2004,
  author = {Geddes, C. G. R. and Toth, Cs. and Van Tilborg, J. and Esarey, E. and Schroeder, C. B. and Bruhwiler, D. and Nieter, C. and Cary, J. and Leemans, W. P.},
  title = {High-Quality Electron Beams from a Laser Wakefield Accelerator Using Plasma-Channel Guiding},
  journal = {Nature},
  volume = {431},
  pages = {538--541},
  year = {2004},
  doi = {10.1038/nature02900}
}

@article{faure_laserplasma_2004,
  author = {Faure, J. and Glinec, Y. and Pukhov, A. and Kiselev, S. and Gordienko, S. and Lefebvre, E. and Rousseau, J.-P. and Burgy, F. and Malka, V.},
  title = {A Laser-Plasma Accelerator Producing Monoenergetic Electron Beams},
  journal = {Nature},
  volume = {431},
  pages = {541--544},
  year = {2004},
  doi = {10.1038/nature02963}
}

@article{blumenfeld_energy_2007,
  author = {Blumenfeld, Ian and Clayton, Christopher E. and Decker, Franz-Josef and Hogan, Mark J. and Huang, Chengkun and Ischebeck, Rasmus and Iverson, Richard and Joshi, Chandrashekhar and Katsouleas, Thomas and Kirby, Neil and Lu, Wei and Marsh, Kenneth A. and Mori, Warren B. and Muggli, Patric and Oz, Erdem and Siemann, Robert H. and Walz, Dieter and Zhou, Miaomiao},
  title = {Energy Doubling of 42 {GeV} Electrons in a Metre-Scale Plasma Wakefield Accelerator},
  journal = {Nature},
  volume = {445},
  pages = {741--744},
  year = {2007},
  doi = {10.1038/nature05538}
}

@article{miao_optical_2020,
  author = {Miao, B. and Feder, L. and Shrock, J. E. and Goffin, A. and Milchberg, H. M.},
  title = {Optical Guiding in Meter-Scale Plasma Waveguides},
  journal = {Phys. Rev. Lett.},
  volume = {125},
  pages = {074801},
  year = {2020},
  doi = {10.1103/PhysRevLett.125.074801}
}

@article{picksley_matched_2024,
  author = {Picksley, A. and Stackhouse, J. and Benedetti, C. and Nakamura, K. and Tsai, H. E. and Li, R. and Miao, B. and Shrock, J. E. and Rockafellow, E. and Milchberg, H. M. and Schroeder, C. B. and van Tilborg, J. and Esarey, E. and Geddes, C. G. R. and Gonsalves, A. J.},
  title = {Matched Guiding and Controlled Injection in Dark-Current-Free, 10-{GeV}-Class, Channel-Guided Laser-Plasma Accelerators},
  journal = {Phys. Rev. Lett.},
  volume = {133},
  pages = {255001},
  year = {2024},
  doi = {10.1103/PhysRevLett.133.255001}
}

@article{miao_benchmarking_2024,
  author = {Miao, B. and Rockafellow, E. and Shrock, J. E. and Hancock, S. W. and Gordon, D. and Milchberg, H. M.},
  title = {Benchmarking of Hydrodynamic Plasma Waveguides for Multi-{GeV} Laser-Driven Electron Acceleration},
  journal = {Phys. Rev. Accel. Beams},
  volume = {27},
  pages = {081302},
  year = {2024},
  doi = {10.1103/PhysRevAccelBeams.27.081302}
}

@article{shrock_guided_2024,
  author = {Shrock, J. E. and Rockafellow, E. and Miao, B. and Le, M. and Hollinger, R. C. and Wang, S. and Gonsalves, A. J. and Picksley, A. and Rocca, J. J. and Milchberg, H. M.},
  title = {Guided Mode Evolution and Ionization Injection in Meter-Scale Multi-{GeV} Laser Wakefield Accelerators},
  journal = {Phys. Rev. Lett.},
  volume = {133},
  pages = {045002},
  year = {2024},
  doi = {10.1103/PhysRevLett.133.045002}
}

@misc{an_situ_2025,
  author = {An, Xiangyan and Chen, Min and Liu, Jianglai and Bai, Zhan and Ji, Liangliang and Sheng, Zhengming and Zhang, Jie},
  title = {In situ axion generation and detection in laser-driven wakefields},
  year = {2025},
  eprint = {2504.12500},
  archivePrefix = {arXiv},
  primaryClass = {physics.plasm-ph},
  doi = {10.48550/arXiv.2504.12500}
}

@article{beyer_light-shining-through-wall_2022,
  author = {Beyer, K. A. and Marocco, G. and Bingham, R. and Gregori, G.},
  title = {Light-shining-through-wall axion detection experiments with a stimulating laser},
  journal = {Phys. Rev. D},
  volume = {105},
  pages = {035031},
  year = {2022},
  doi = {10.1103/PhysRevD.105.035031}
}

@misc{AxionLimits,
  author = {O'Hare, Ciaran},
  title = {cajohare/{AxionLimits}: {AxionLimits}},
  publisher = {Zenodo},
  version = {v1.0},
  year = {2020},
  doi = {10.5281/zenodo.3932430},
  howpublished = {\url{https://cajohare.github.io/AxionLimits/}}
}

\end{document}